\documentclass[aps,twocolumn,showpacs,preprintnumbers,amsmath,amssymb,
endfloats]{revtex4}
\pdfoutput=1
\usepackage{graphicx}

\begin{document}

\title{Surface plasmon resonances of metallic nanostars/nanoflowers for surface-enhanced Raman scattering}

\author{V. Giannini}
\email[Present address: Center for Nanophotonics, FOM Institute AMOLF, c/o Philips Research Laboratories, High Tech Campus 4, 5656 AE Eindhoven, The Netherlands.]{}
%Electronic mail: ]{vingenzino@iem.cfmac.csic.es}
\author{R. Rodr\'{\i}guez-Oliveros}
%\email[Electronic mail: ]{rogelio@iem.cfmac.csic.es}
\author{J. A. S{\'a}nchez-Gil}
\email[Electronic mail: ]{j.sanchez@iem.cfmac.csic.es}
\affiliation{Instituto de Estructura de la Materia,
Consejo Superior de Investigaciones Cient{\'\i}ficas, Serrano 121,
28006 Madrid, Spain.}
\date{\today}
\begin{abstract}

We investigate theoretically the optical properties associated to plasmon resonances of metal nanowires with cross section given by low-order Chebyshev nanoparticles (like rounded-tip nanostars or nanoflowers). The impact of the nanoflower shape is analyzed for varying symmetry and deformation parameter through the spectral dependence of resonances, and their corresponding near field distributions. Large field  intensity enhancements are obtained at the gaps between petals, apart from the tips themselves, which make these nanostars/nanoflowers specially suitable to host molecules for SERS sensing applications. 

\end{abstract}

%\ocis{}
%\pacs{78.67.bf, 42.25.Fx, 73.20.mf}

\maketitle

\section*{Introduction}

Surface-enhanced Raman scattering (SERS)
has attracted a renewed interest in the last decade, due to both single molecule detection and  advances in nanofabrication \cite{SERS_AC05}. Bear in mind that SERS spectra provide (bio)molecular fingerprints that renders SERS spectroscopy an extremely powerful sensing technique \cite{SERS_AC05,VanDuynesensor05}. The large electromagnetic field enhancements thus required (Raman scattering cross sections being remarkably low) are typically achieved by using  colloidal aggregates of noble metal nanoparticles (NPs) as SERS substrates \cite{Aroca_SERS}. Hot spots (HS) occur at given intersticies in between NPs, thereby enhancing the Raman scattering of molecules adsorbed nearby. 
On the other hand, to avoid issues related to reproducibility and control of colloidal NPs, new nanofabrication techniques have been proposed based on self-assembling layers or on nanolithography \cite{SERS_AC05}; in particular, the latter approach has given rise to the so-called dimer nanoantennas \cite{Antenna_science,Moerner08_SERSant,NL07,OL08}.

Surface plasmon resonances (SPRs) underly the occurrence of HS \cite{K-V}. Nonetheless, strong coupling between NPs is needed to produce HS yielding large enough field enhancements \cite{Aizpurua_PRE00,Schatz04,LeRu06}. To this end, SERS substrates based on NP aggregates (or at least NP dimers) are required, which in most cases complicate fabrication procedures \cite{SERS_AC05,Aroca_SERS}. Alternatively, strong SPRs with significant HS could be achieved on single NPs of complex shape, as expected from symmetry breaking arguments. 
In recent years,  the light scattering properties of nanoparticles with star (or flower) shape have been experimentally studied \cite{Nehl06_star,Liz08,ACS08_flower}; indeed,  nanostar/nanoflower NPs have been shown suitable as SERS substrates \cite{JPC08_star,JRS09_star,ACS08_flower,APL09_star}. From the theoretical standpoint, recent SPR calculations indicate that HS may take place at the star tips \cite{Nord07_star,Liz08}, but further work is no doubt needed to shed light on their optical properties, due to the variety of NP shapes experimentally produced.

In this work we investigate theoretically the SPRs of silver nanowires with cross section given by low-order Chebyshev nanoparticles with large deformations, which resemble either nanostars with round tips or nanoflowers, with different number of tips/petals. We will simply refer to them as nanoflowers throughout the rest of this work for the sake of simplicity. Incidentally, light scattering from  dielectric Chebishev particles have been investigated in connection with remote sensing and related (cf. i.e. Ref. \onlinecite{Mish-book}).  We will show that metal nanoflowers exhibit a complex SPR spectrum; calculated near-field distributions at resonant frequencies reveal that HS with large field enhancement factors are found at the gaps in between petals. 

\section*{Theory}

\begin{figure}
\centering
\includegraphics[width=\columnwidth]{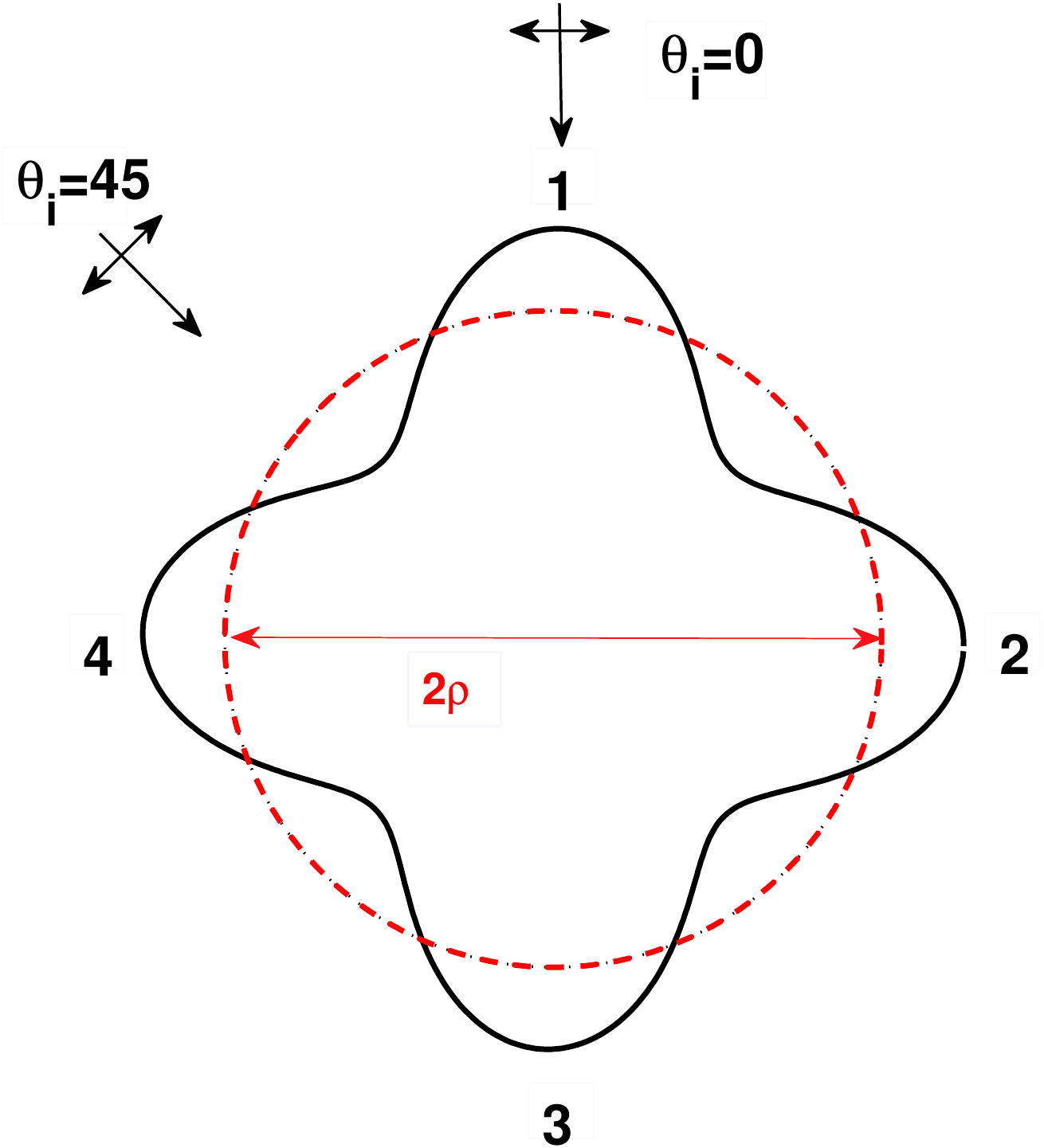}
\caption{\footnotesize{Illustration of the star/flower shape given by a Chebyshev particle: the profile is a sum of a circle equation (radius $\rho=40$ nm) and a cosine function with four periods (deformation parameter $\beta=10$ nm).}}
\label{star_illu}
\end{figure}

The scattering configuration we consider is shown in Fig. \ref{star_illu}.
We study the light scattering from two-dimensional Ag nanostars/nanoflowers, which exhibit a
translational symmetry along one direction, such as nanowires. 
We have simulated the star/flower shape as the sum of a circle (radius $\rho$) and 
a cosine function with $n$ periods of amplitude $2\beta$ (Chebyshev NPs), $\beta$ being the deformation parameter; the corresponding surface profile  of the flower with $n$ petals  in the $xz$ plane is represented  by the following parametric curve in polar coordinates:
\begin{eqnarray}
\rho_n\equiv \rho(1+\beta\cos n\phi), \phi\in[0,2\pi) , \beta< 1
\end{eqnarray}
so that the continuous cartesian vector-valued function $\textbf{R})$ is defined as:
\begin{eqnarray}
\textbf{R}_n(\rho,\phi)\equiv \big(-\rho\sin\phi[1+\beta\cos(n\phi)],\\\nonumber
                          -\rho\cos\phi[1+\beta\cos(n\phi)]\big).
\end{eqnarray}
 The isotropic, homogeneous,
frequency-dependent dielectric function for bulk Ag is taken from Ref.
\onlinecite{J-C}; neglecting electron surface scattering effects is justified
as long as the radius of curvature of flower petals exceeds the electron mean free
path. 

The nanoflowers are illuminated in the plane of the figure with a
$p$-polarized, monochromatic plane electromagnetic wave of frequency $\omega$ at
an angle $\theta_{0}$ with the $z$ axis. The polarization is defined as shown in
Fig. \ref{star_illu}: the magnetic field is perpendicular to the $xz$ plane 
(transverse magnetic polarization). The restriction to two-dimensional systems
and linear polarization has the advantage that notably simplifies the formulation, 
thus reducing the initial  three-dimensional vectorial wave equation to a 
two-dimensional scalar one, where the EM field in this polarization is entirely 
described by the $y$-component of  the magnetic field $H_y=H(\mathbf{r})$. 
The method of calculation is based on the surface integral equations  resulting 
from the application of Green's second integral theorem adapted to
parametric coordinates \cite{JOSAA07}, which is exact from the
standpoint of classical electrodynamics (fully accounting for multipoles and
retardation). In this manner, rigorous calculations are carried out for the
spectral dependence of the scattering cross sections (SCS), from which plasmon
resonances are identified; relevant plasmon resonances are subsequently
characterized through detailed calculations of near-field
distributions.

\section*{Results and discussion}

\begin{figure}
\centering
\includegraphics[width=\columnwidth]{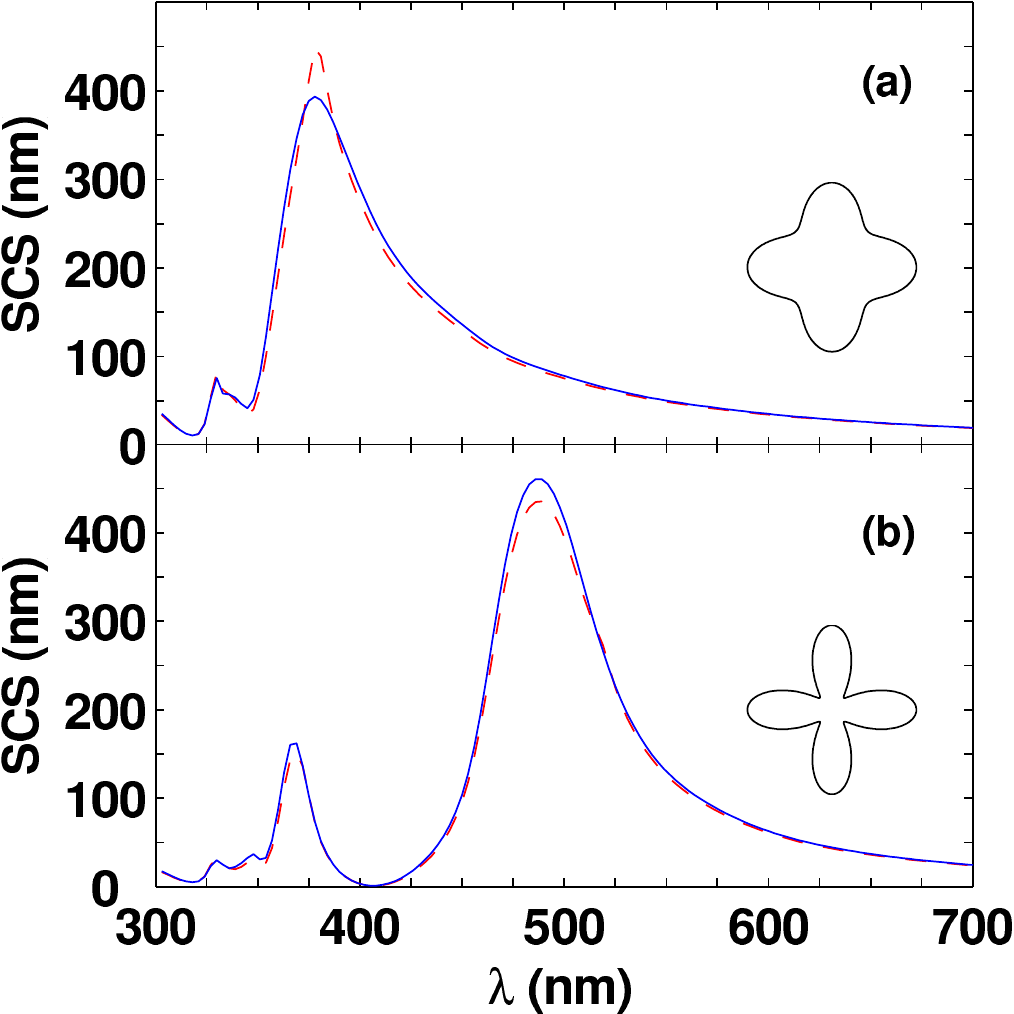}
\caption{\footnotesize{Scattering cross section ($p$ polarization) for Ag 
nanoflowers with 4 petals at two different illumination directions (as defined in Fig.~\protect{\ref{star_illu}} impinging on a tip ($\theta_i = 0^{\circ}$, blue solid curves) or on a dip ($\theta_i = 45^{\circ}$, red dashed curves). Two nanoflowers are considered:
(a) $\rho=40$ nm and $\beta=1/4$;
(b) $\rho=30$ nm and $\beta=2/3$.}}
\label{scs_star4}
\end{figure}

%
% SCS 4
%
In figure \ref{scs_star4}, the SCS of two 4-petal Ag nanoflowers are shown for two different symmetry directions of the incident field (on the upper tip and on a valley, as depicted in figure \ref{star_illu}). The parameters of the two nanoflowers are:  $\rho=40$ nm and $\beta=1/4$, and   $\rho=30$ nm and $\beta=2/3$. They are chosen so that the resulting maximum radius is identical ($R=\rho(1+\beta)=50$ nm), but the inner radius is different ($r=\rho(1-\beta)=30$ and 10 nm, respectively). Thus we can observe the variations of the SPRs from a nearly circular NP to a strongly modified nanoflower with long petals. Expectedly, the SCS of the nearly circular nanostar shown in Fig. \ref{scs_star4}b exhibits for both incident fields a strong SPR at $\lambda=376$ nm, along with the weaker bulk plasmon resonance at $\lambda=329$ nm. Both are closely related to the SPRs of a circular cylinder \cite{JOSAA07}, the strong dipole-like SPR being slightly red-shifted. In contrast, two strong SPRs are observed in the SCS of the nanoflower shown in Fig. \ref{scs_star4}a: at $\lambda=487$  nm and at $\lambda=369$ nm. Interestingly, the SCS for the two illumination directions are also quite similar in this case. 

\begin{figure}
\centering
\includegraphics[width=\columnwidth]{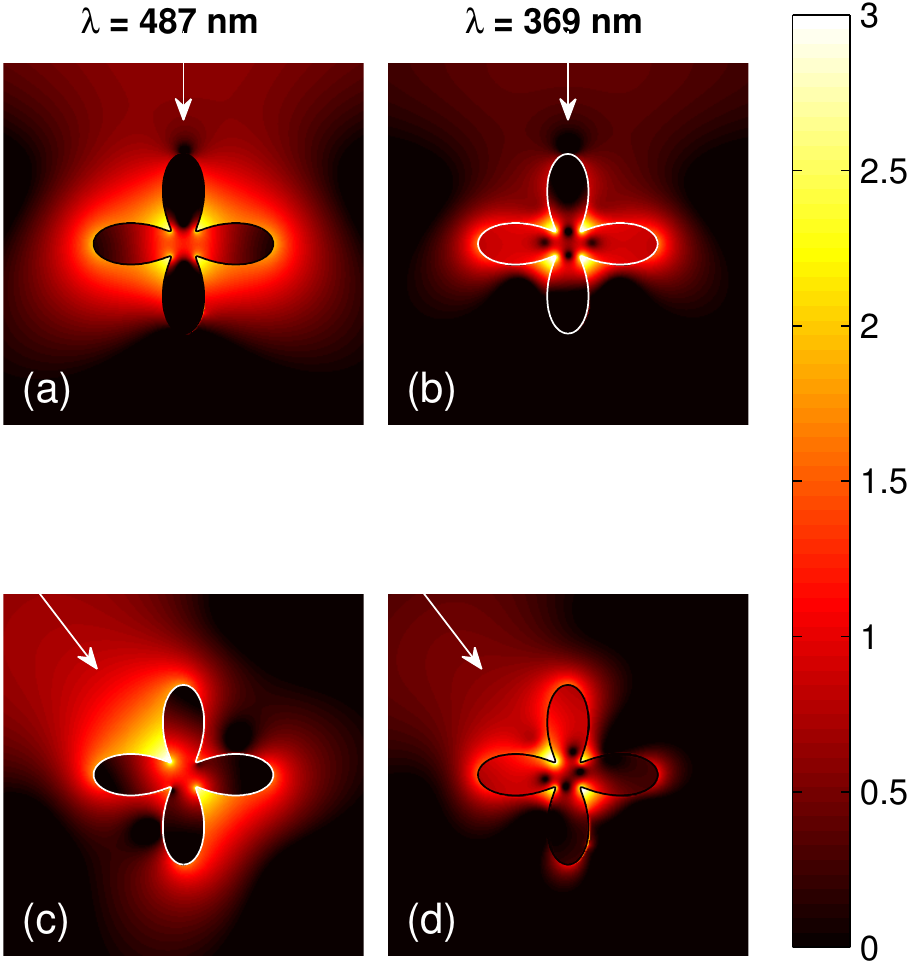}
\caption{\footnotesize{Near-field distributions of the electric field intensities
in a log$_{10}$-scale for the Ag 4-petal nanoflower with mean radius $\rho=30$ nm and deformation parameter  $\beta=2/3$ at the two main SPRs at $\lambda=487$ nm (a,c) and  at $\lambda=369$ nm (b,d):
(a,b) $\theta_i = 0^{\circ}$; (c,d) $\theta_i = 45^{\circ}$.}}
\label{nf_star4}
\end{figure}

%
% NF 4
%
In order to identify the corresponding SPRs of the nanoflower, the near-field intensity distributions at the main resonances are shown in Fig.~\ref{nf_star4}. Let us first focus on the main resonance at $\lambda=487$ nm shown in Figs.~\ref{nf_star4}a and~\ref{nf_star4}c. Despite the SCS at $\theta_i=0^{\circ},45^{\circ}$ are barely distinguishable, the near-field patterns are clearly different. Both exhibit a dipole-like behavior, but the illumination direction seems to impose a dipole-electric field distribution oriented along the polarization direction; namely, along the horizontal petals for $\theta_i=0^{\circ}$, and across opposite walls between adjacent petals for $\theta_i=45^{\circ}$. In fact, HS with large field intensity enhancements (about three orders of magnitude) are located at all four interstitial sites for $\theta_i=0^{\circ}$, unlike those HS for  $\theta_i=45^{\circ}$, that only appear at the two dips between petals along the illumination direction. In either case, this dipolar character indicates that this main SPR stems from the dipole SPR of a single cylinder, largely red-shifted due to the complex shape of the 4-tip nanoflower. 

%
% NF 4
%
The near field distributions of the weaker SPR at $\lambda=369$ nm are shown in  Figs.~\ref{nf_star4}b and~\ref{nf_star4}d. It is evident from Fig.~\ref{nf_star4}b that  this high-energy SPR exhibits a more complex, $\sim$4-fold symmetry pattern, but not exactly quadrupolar, as will be shown below. Again, the near-field distributions is different depending on the illumination direction unlike the resonance SCS spectra, although a deformed 4-fold pattern is also observed for $\theta_i=45^{\circ}$ in Fig.~\ref{nf_star4}d. HS are also found, however weaker than those at the main resonance, distributed either uniformly  among the four  interstitial sites for  $\theta_i=0^{\circ}$ or at the two interstitial sites along the illumination for  $\theta_i=45^{\circ}$.
Incidentally, the existence of multiple SPRs yielding HS at overlapping regions make nanoflowers also suitable for the two-fold (excitation and emission) enhancement of fluorescence, as predicted
for nanotrimers \cite{OL08}.

\begin{figure}
\centering
\includegraphics[width=\columnwidth]{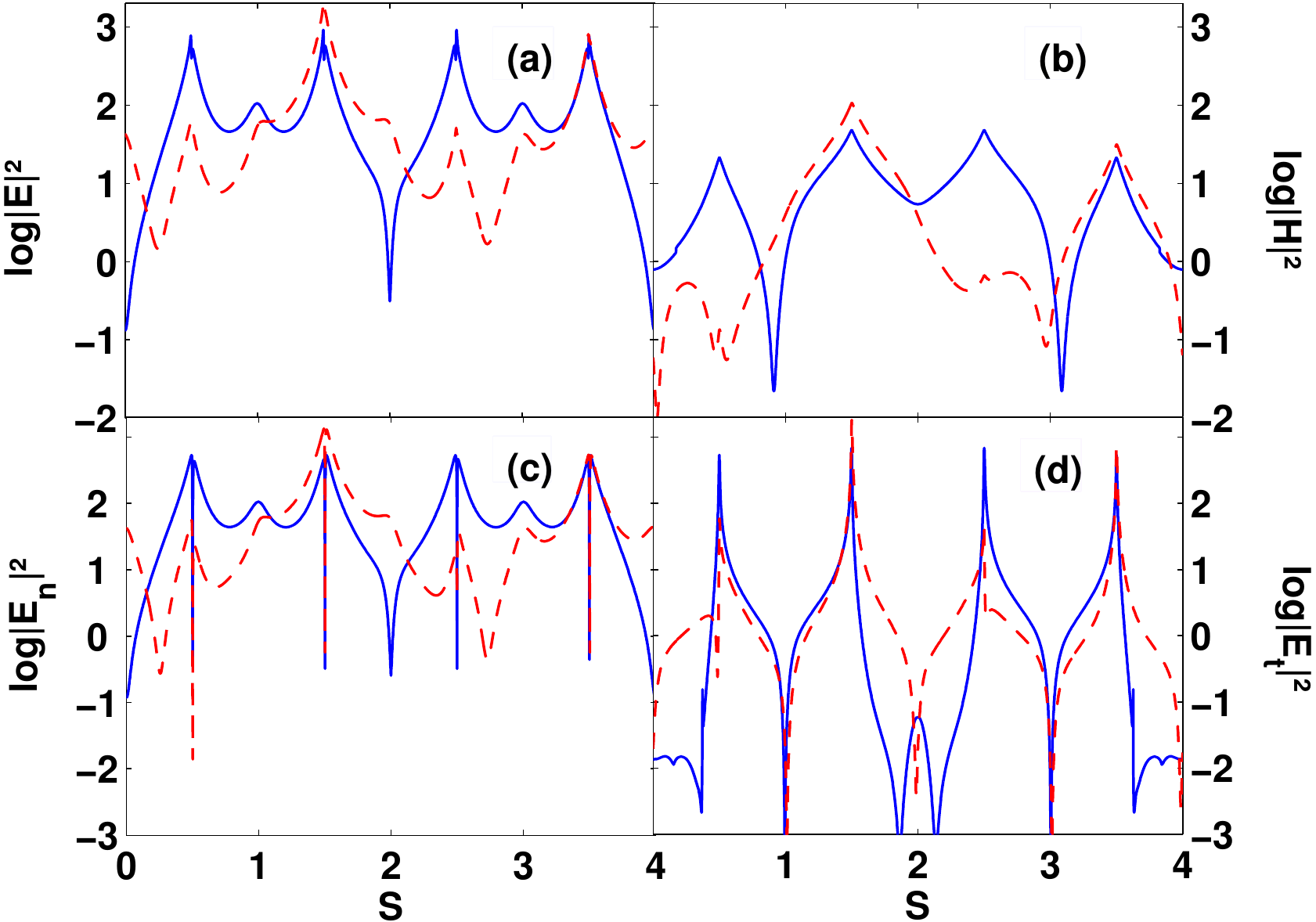}
\caption{\footnotesize{Surface-field intensities on the 4-petal 
Ag nanoflower with $\rho=30$ nm and $\beta=2/3$ at the main SPR ($\lambda=487$ nm) 
for two illumination directions (blue solid curve $\theta_i=0^{\circ}$, 
red dashed red curve $\theta_i=45^{\circ}$): (a) Electric field intensity; 
(b) Magnetic field intensity; (c) Normal component of the electric field intensity; and
(d) Tangential component of the electric field intensity. The integer values of the $s$ parameter
variable in the abscissas correspond to the four petal tips, whereas interstitial gaps
between petals lie at half-integer values (see Fig. \ref{star_illu}).}}
\label{sf_star4}
\end{figure}

%
% SF 4
%
To get further insight into the HS at the main resonance (see Fig.~\ref{sf_star4}a), the corresponding surface (electric and magnetic) field components are presented in  Figs.~\ref{sf_star4}. The abscissae axis in parametric coordinates is such that integer values (from $s=1$ to 4) correspond to flower tips, so that half-integer values identify interstitial gaps. First of all, note the electric field intensity (Fig.~\ref{sf_star4}a) exhibits maxima at both tips and intersticies, whereas the magnetic field intensity (Fig.~\ref{sf_star4}a) presents maxima only on tips. Since we are mainly concerned about electric field intensity enhancements for SERS applications, let us examine these HS in detail. At tips, the contribution to the HS stems from the normal component of the electric field intensity (Fig.~\ref{sf_star4}c). However, when the incident plane wave impinges perpendicular to the tip (see Figs.~\ref{sf_star4}a and~\ref{sf_star4}c $\theta_i=0^{\circ}$ at $n=0,2$), instead of a HS, the electric field nearly vanishes. Interestingly, such surface field patterns resemble those of the shape resonances associated to localized surface-plasmon polaritons with monopole-like behavior found at peaks on rough metal surfaces \cite{PRB03spp}, which cannot be excited at normal incidence with electric field polarization tangential to the tip/peak. On the other hand, the surface field pattern at the interstitial gaps has also a normal component (see Fig.~\ref{sf_star4}c), except at the very edge of the dip, where a strong contribution from the tangential component occurs (see Fig.~\ref{sf_star4}d). This behavior is also found on randomly metal surfaces, associated to localized surface-plasmon polaritons with dipole-like resonances \cite{PRB03spp}.
Overall, enhancements factors of the electric field intensity about two orders of magnitude are found at both tips and dips (see Fig.~\ref{sf_star4}a); nonetheless, the HS area as inferred from the near-field patterns is substantially larger at the interstitial gaps between tips (recall Figs.~\ref{nf_star4}a and~\ref{nf_star4}c). SERS enhancement factors up to $10^{7}$ are indeed found in the most favorable configuration (in the gap directly illuminated by the incident plane wave at $\theta_i=45^{\circ}$), although more HS occur at $\theta_i=0^{\circ}$.

\begin{figure}
\centering
\includegraphics[width=\columnwidth]{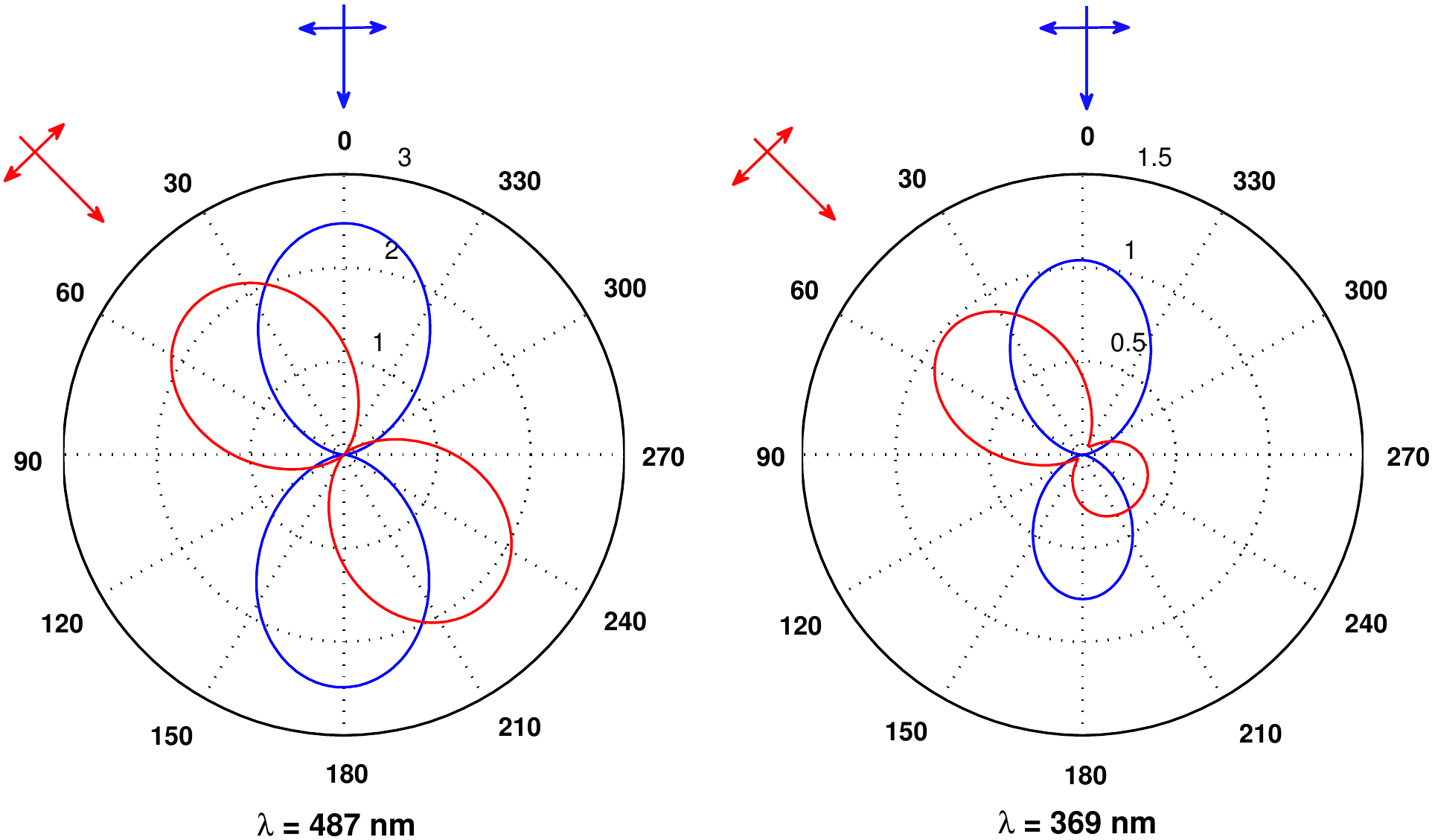}
\caption{\footnotesize{Angular distribution of the far-field intensity scattered from 
the 4-petal Ag nanoflower with $\rho=30$ nm and $\beta=2/3$ for two illumination 
directions (blue solid curve $\theta_i=0^{\circ}$, red dashed red curve $\theta_i=45^{\circ}$) 
at the two main SPRs: (a) $\lambda=487$ nm and (b) $\lambda=369$ nm.
$\theta = 0$ is the forward direction.}}
\label{ff_star4}
\end{figure}

%
% FF 4
%
For the sake of completeness, the far-field patterns yielded by the 4-petal nanoflower at the two main resonances described in Figs.~\ref{nf_star4}a and~\ref{nf_star4}c is shown in Fig.~\ref{ff_star4}. The dipolar character of the main resonance at $\lambda=487$ nm is confirmed by the behavior of the far-field pattern (see Fig.~\ref{ff_star4}a), which accurately resembles that produced by oscillating dipoled oriented along the direction of the incident polarization (vertical for $\theta_i=0^{\circ}$ and $45^{\circ}$-tilted counterclockwise for $\theta_i=45^{\circ}$). At the higher-frequency SPR at $\lambda=367$ nm (Fig.~\ref{ff_star4}b), the far-field distributions are mostly dipolar, but with asymmetric lobes. This indicates that the 4-fold symmetry observed in the near-field patterns in Figs.~\ref{nf_star4}b and~\ref{nf_star4}d does not lead to quadrupolar charge oscillation, but to a predominant dipole oscillation again oriented along the direction of the incident polarization.

\begin{figure}
\centering
\includegraphics[width=\columnwidth]{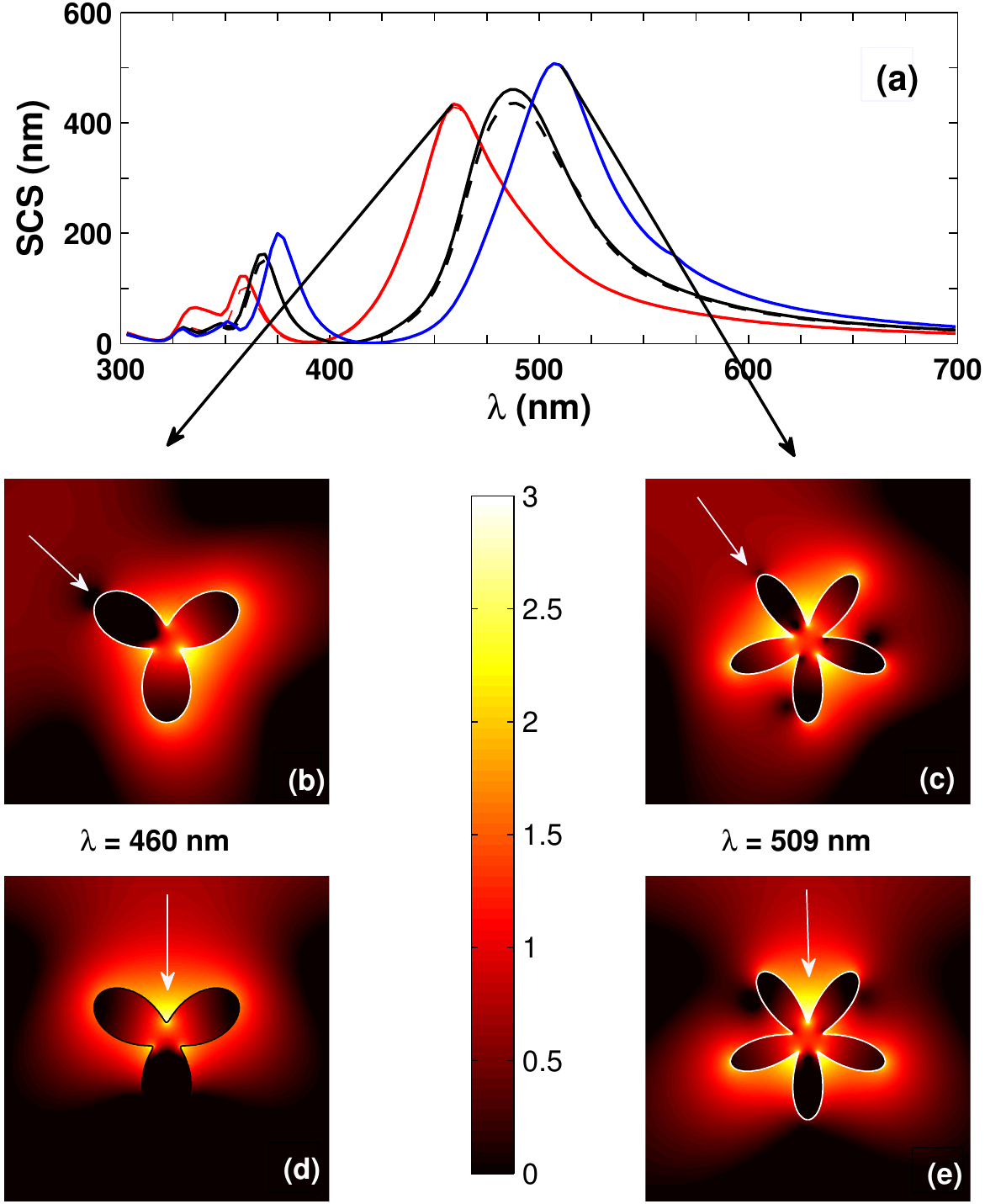}

\caption{\footnotesize{(a) Scattering cross section ($p$ polarization) for Ag nanoflowers with 3 and 4 petals both with the same parameters as that in Fig.~\protect{\ref{scs_star4}} ($\rho=30$ nm and amplitude $\beta=2/3$) at two different illumination directions impinging on a dip ($\theta_i = 0^{\circ}$, dashed curves) or on a tip (solid curves, $\theta_i = 60^{\circ}$ for the 3-tip nanoflower and  $\theta_i = 72^{\circ}$ for the 5-tip nanoflower). Near-field distributions (electric field intensities in a log$_{10}$-scale) at the main SPRs ($\lambda=460$ nm for the 3-petal nanoflower and $\lambda=509$ nm for the 5-petal nanoflower) are shown:  (b)  $\theta_i = 60^{\circ}$;  (c) $\theta_i = 36^{\circ}$; and (b,d) $\theta_i = 0^{\circ}$.}}
\label{scsnf_star35}
\end{figure}

%
% SCS,NF 3-5
%
Finally, we investigate HS at nanoflowers with a different number of petals. In particular, odd numbers $n=3,5$ are chosen so that opposite petals are not aligned, keeping both radius and deformation parameter fixed ($\rho=30$ nm, $\beta=2/3$). The resulting SCS are shown in Fig.~\ref{scsnf_star35}a. First, note that the qualitative behavior is similar to that of the 4-petal nanoflower shown in Fig.~\ref{scs_star4}b. A strong, red-shifted SPR with a weaker one at higher energies. Indeed, both resonances are red-shifted (respectively, blue-shifted) with respect to those of the 4-petal nanoflower for larger (respectively, smaller) number of petals. In addition, the near-field patterns associated with all resonances behave qualitatively similar to those of the two main resonances of the 4-petal nanoflower shown in Fig.~\ref{nf_star4}. Here only those at the main SPRs  are shown in Figs.~\ref{scsnf_star35}b and~\ref{scsnf_star35}d for the 3-petal nanoflower ($\lambda=460$ nm), and in Figs.~\ref{scsnf_star35}c and~\ref{scsnf_star35}e for the 5-petal nanoflower ($\lambda=509$ nm). Despite the different number of petals, basically a dipolar character is preserved in the near-field distributions, with the orientation determined by the incident polarization, leading to HS at the interstitial gaps along the illumination direction, and a dark pattern at the illuminated petal. For plane wave illumination at a dip (see Figs.~\ref{scsnf_star35}d and~\ref{scsnf_star35}e), the dipolar behavior is preserved, so that the dark petal is now that opposite to the illumination direction,  with  HS again at dips along the illumination direction. SERS enhancement factors of the order of $10^6$ are also obtained. 

Moreover, if the near-field patterns at $\theta_i=0^{\circ}$ in Figs.~\ref{scsnf_star35}b and~\ref{scsnf_star35}c are rotated by certain angle $\alpha$, both coincide with those in Figs.~\ref{scsnf_star35}d and~\ref{scsnf_star35}e, respectively: namely, by (counterclockwise) $\alpha=180^{\circ}[1-1/(2n)]$ with $n=3$ for Fig.~\ref{scsnf_star35}b
and $n=5$ for Fig.~\ref{scsnf_star35}c. Basically, this means that the same near-field intensity pattern is retrieved if the nanoflower is illuminated along either a petal tip  or its opposite dip. Intuitively, this stems from the mirror symmetry of this dipolar SPR with respect to a line parallel to the incident direction crossing the nanoflower center. Indeed, note that the SCS at the main SPRs in Fig.~\ref{scsnf_star35}a are identical. This symmetry considerations only hold for nanoflowers with odd numer of petals. In the case of an even number of petals, the main SPRs differ at different angles (illuminating either tip or dip, see Figs.~\ref{nf_star4}a and~\ref{nf_star4}c), though each of them exhibits symmetric near-field patterns with respect to the axis along the incident direction); thus the associated SCS slightly differ too (see Fig.~\ref{scs_star4}b). Furthermore, such symmetry argument no longer holds for the weaker, high-energy SPRs, even for odd-$n$ nanoflowers (not shown here). Surface charge oscillations will be investigated in detail elsewhere.

\section*{Conclusions}

In this work we have investigated theoretically the SPRs of silver nanowires with
cross section given by low-order Chebyshev nanoparticles with large deformation parameter. They resemble either  nanostars with round tips or nanoflowers, with different number of tips/petals.
We have shown that such metal nanoflowers exhibit a complex SPR spectrum, with a main SPR red-shifted with respect to that of a circular wire, and a weaker SPR appearing in the blue part of the spectra. Calculated near-field distributions for the main SPRs reveal a dipolar behavior (confirmed by the far-field pattern), with strong HS at both tips and dips of the petals. However, apart from symmetry considerations, HS occurring at the intersticies in between petals occupy larger areas and yield larger field enhancement factors. In fact, SERS enhancement factors up to $10^7$ are found at various HS for nanoflowers with different number of petals. Thus nanoflowers (or nanostars with rounded tips)  manifest themselves as excellent candidates (monodisperse, single nanoparticle substrates)  for SERS  sensing applications, where the interstitial gaps between petals may both host the (bio)molecule to be detected and yield the required electromagnetic field enhancement.

\section*{Acknowledments}

This work was supported by the Spanish MICINN (grants
FIS2006-07894 and Consolider-Ingenio 2010 \textit{EMET} CSD2008-00066) and the Comunidad de
Madrid (grant \textit{MICROSERES} S-0505/TIC-0191).

%
%\bibliography{mybib}

\begin{thebibliography}{22}
\expandafter\ifx\csname natexlab\endcsname\relax\def\natexlab#1{#1}\fi
\expandafter\ifx\csname bibnamefont\endcsname\relax
  \def\bibnamefont#1{#1}\fi
\expandafter\ifx\csname bibfnamefont\endcsname\relax
  \def\bibfnamefont#1{#1}\fi
\expandafter\ifx\csname citenamefont\endcsname\relax
  \def\citenamefont#1{#1}\fi
\expandafter\ifx\csname url\endcsname\relax
  \def\url#1{\texttt{#1}}\fi
\expandafter\ifx\csname urlprefix\endcsname\relax\def\urlprefix{URL }\fi
\providecommand{\bibinfo}[2]{#2}
\providecommand{\eprint}[2][]{\url{#2}}

\bibitem{SERS_AC05}
\bibinfo{author}{\bibfnamefont{C.~L.} \bibnamefont{Haynes}},
  \bibinfo{author}{\bibfnamefont{A.~D.} \bibnamefont{McFarland}},
  \bibnamefont{and} \bibinfo{author}{\bibfnamefont{R.~P.~V.}
  \bibnamefont{Duyne}}, \bibinfo{journal}{Anal. Chem.}
  \textbf{\bibinfo{volume}{77}}, \bibinfo{pages}{338A} (\bibinfo{year}{2005}).

\bibitem{VanDuynesensor05}
\bibinfo{author}{\bibfnamefont{C.~R.} \bibnamefont{Yonzon}},
  \bibinfo{author}{\bibfnamefont{D.~A.} \bibnamefont{Stuart}},
  \bibinfo{author}{\bibfnamefont{X.}~\bibnamefont{Zhang}},
  \bibinfo{author}{\bibfnamefont{A.~D.} \bibnamefont{{McFarland}}},
  \bibinfo{author}{\bibfnamefont{C.~L.} \bibnamefont{Haynes}},
  \bibnamefont{and} \bibinfo{author}{\bibfnamefont{R.~P.} \bibnamefont{{Van
  Duyne}}}, \bibinfo{journal}{Talanta} \textbf{\bibinfo{volume}{67}},
  \bibinfo{pages}{438} (\bibinfo{year}{2005}).

\bibitem{Aroca_SERS}
\bibinfo{author}{\bibfnamefont{R.}~\bibnamefont{Aroca}},
  \emph{\bibinfo{title}{Surface-Enhanced Vibrational Spectroscopy}}
  (\bibinfo{publisher}{John Wiley, New York}, \bibinfo{year}{2006}).

\bibitem{Antenna_science}
\bibinfo{author}{\bibfnamefont{P.}~\bibnamefont{M{\"u}hlschlegel}},
  \bibinfo{author}{\bibfnamefont{H.-J.} \bibnamefont{Eisler}},
  \bibinfo{author}{\bibfnamefont{O.~J.~F.} \bibnamefont{Martin}},
  \bibinfo{author}{\bibfnamefont{B.}~\bibnamefont{Hecht}}, \bibnamefont{and}
  \bibinfo{author}{\bibfnamefont{D.~W.} \bibnamefont{Pohl}},
  \bibinfo{journal}{Science} \textbf{\bibinfo{volume}{308}},
  \bibinfo{pages}{1607} (\bibinfo{year}{2005}).

\bibitem{Moerner08_SERSant}
\bibinfo{author}{\bibfnamefont{F.}~\bibnamefont{J{\"a}ckel}},
  \bibinfo{author}{\bibfnamefont{A.~A.} \bibnamefont{Kinkhabwala}},
  \bibnamefont{and} \bibinfo{author}{\bibfnamefont{W.~E.}
  \bibnamefont{Moerner}}, \bibinfo{journal}{Chem. Phys. Lett.}
  \textbf{\bibinfo{volume}{446}}, \bibinfo{pages}{339} (\bibinfo{year}{2007}).

\bibitem{NL07}
\bibinfo{author}{\bibfnamefont{O.~L.} \bibnamefont{Muskens}},
  \bibinfo{author}{\bibfnamefont{V.}~\bibnamefont{Giannini}},
  \bibinfo{author}{\bibfnamefont{J.~A.} \bibnamefont{S{\'a}nchez-Gil}},
  \bibnamefont{and} \bibinfo{author}{\bibfnamefont{J.~G.} \bibnamefont{Rivas}},
  \bibinfo{journal}{Nano Lett.} \textbf{\bibinfo{volume}{7}},
  \bibinfo{pages}{2871} (\bibinfo{year}{2007}).

\bibitem{OL08}
\bibinfo{author}{\bibfnamefont{V.}~\bibnamefont{Giannini}} \bibnamefont{and}
  \bibinfo{author}{\bibfnamefont{J.~A.} \bibnamefont{S{\'a}nchez-Gil}},
  \bibinfo{journal}{Opt. Lett.} \textbf{\bibinfo{volume}{33}},
  \bibinfo{pages}{899} (\bibinfo{year}{2008}).

\bibitem{K-V}
\bibinfo{author}{\bibfnamefont{U.}~\bibnamefont{Kreibig}} \bibnamefont{and}
  \bibinfo{author}{\bibfnamefont{M.}~\bibnamefont{Vollmer}},
  \emph{\bibinfo{title}{Optical Properties of Metal Clusters}}
  (\bibinfo{publisher}{Springer, Berlin}, \bibinfo{year}{1995}).

\bibitem{Aizpurua_PRE00}
\bibinfo{author}{\bibfnamefont{H.}~\bibnamefont{Xu}},
  \bibinfo{author}{\bibfnamefont{J.}~\bibnamefont{Aizpurua}},
  \bibinfo{author}{\bibfnamefont{M.}~\bibnamefont{K{\"a}ll}}, \bibnamefont{and}
  \bibinfo{author}{\bibfnamefont{P.}~\bibnamefont{Apell}},
  \bibinfo{journal}{Phys. Rev. E} \textbf{\bibinfo{volume}{62}},
  \bibinfo{pages}{4318} (\bibinfo{year}{2000}).

\bibitem{Schatz04}
\bibinfo{author}{\bibfnamefont{E.}~\bibnamefont{Hao}} \bibnamefont{and}
  \bibinfo{author}{\bibfnamefont{G.~C.} \bibnamefont{Schatz}},
  \bibinfo{journal}{J. Chem Phys.} \textbf{\bibinfo{volume}{120}},
  \bibinfo{pages}{357} (\bibinfo{year}{2004}).

\bibitem{LeRu06}
\bibinfo{author}{\bibfnamefont{E.}~\bibnamefont{Le~Ru}},
  \bibinfo{author}{\bibfnamefont{P.~G.} \bibnamefont{Etchegoin}},
  \bibnamefont{and} \bibinfo{author}{\bibfnamefont{M.~J.} \bibnamefont{Meyer}},
  \bibinfo{journal}{J. Chem. Phys.} \textbf{\bibinfo{volume}{125}},
  \bibinfo{pages}{204701} (\bibinfo{year}{2006}).

\bibitem{Nehl06_star}
\bibinfo{author}{\bibfnamefont{C.~L.} \bibnamefont{Nehl}},
  \bibinfo{author}{\bibfnamefont{H.}~\bibnamefont{Liao}}, \bibnamefont{and}
  \bibinfo{author}{\bibfnamefont{H.}~\bibnamefont{Hafner}},
  \bibinfo{journal}{Nano Lett.} \textbf{\bibinfo{volume}{6}},
  \bibinfo{pages}{683} (\bibinfo{year}{2006}).

\bibitem{Liz08}
\bibinfo{author}{\bibfnamefont{P.}~\bibnamefont{Kumar}},
  \bibinfo{author}{\bibfnamefont{I.}~\bibnamefont{Pastoriza-Santos}},
  \bibinfo{author}{\bibfnamefont{B.}~\bibnamefont{Rodr{\'i}gues-Gonz{\'a}lez}},
  \bibinfo{author}{\bibfnamefont{F.~J.} \bibnamefont{Garc{\'i}a-Abajo}},
  \bibnamefont{and} \bibinfo{author}{\bibfnamefont{L.~M.}
  \bibnamefont{Liz-Marz{\'a}n}}, \bibinfo{journal}{Nanotechnology}
  \textbf{\bibinfo{volume}{19}}, \bibinfo{pages}{015606}
  (\bibinfo{year}{2008}).

\bibitem{ACS08_flower}
\bibinfo{author}{\bibfnamefont{J.}~\bibnamefont{Xie}},
  \bibinfo{author}{\bibfnamefont{Q.}~\bibnamefont{Zhang}},
  \bibinfo{author}{\bibfnamefont{J.~Y.} \bibnamefont{Lee}}, \bibnamefont{and}
  \bibinfo{author}{\bibfnamefont{D.~I.~C.} \bibnamefont{Wang}},
  \bibinfo{journal}{ACS Nano} \textbf{\bibinfo{volume}{2}},
  \bibinfo{pages}{2473} (2008).

\bibitem{JPC08_star}
\bibinfo{author}{\bibfnamefont{C.~G.} \bibnamefont{Khoury}} \bibnamefont{and}
  \bibinfo{author}{\bibfnamefont{T.~J.} \bibnamefont{Vo-Dinh}},
  \bibinfo{journal}{J. Phys. Chem. C} \textbf{\bibinfo{volume}{112}},
  \bibinfo{pages}{18849} (\bibinfo{year}{2008}).

\bibitem{JRS09_star}
\bibinfo{author}{\bibfnamefont{E.~N.} \bibnamefont{Esenturk}} \bibnamefont{and}
  \bibinfo{author}{\bibfnamefont{A.~R.} \bibnamefont{HightWalker}},
  \bibinfo{journal}{J. Raman Spectrosc.} \textbf{\bibinfo{volume}{40}},
  \bibinfo{pages}{86} (\bibinfo{year}{2009}).

\bibitem{APL09_star}
\bibinfo{author}{\bibfnamefont{C.}~\bibnamefont{Hrelescu}},
  \bibinfo{author}{\bibfnamefont{T.~K.} \bibnamefont{Sau}},
  \bibinfo{author}{\bibfnamefont{A.~L.} \bibnamefont{Rogach}},
  \bibinfo{author}{\bibfnamefont{F.}~\bibnamefont{Jackel}}, \bibnamefont{and}
  \bibinfo{author}{\bibfnamefont{J.}~\bibnamefont{Feldmann}},
  \bibinfo{journal}{Appl. Phys. Lett.} \textbf{\bibinfo{volume}{94}},
  \bibinfo{eid}{153113} (\bibinfo{year}{2009}).

\bibitem{Nord07_star}
\bibinfo{author}{\bibfnamefont{F.}~\bibnamefont{Hao}},
  \bibinfo{author}{\bibfnamefont{C.~L.} \bibnamefont{Nehl}},
  \bibinfo{author}{\bibfnamefont{J.~H.} \bibnamefont{Hafner}},
  \bibnamefont{and}
  \bibinfo{author}{\bibfnamefont{P.}~\bibnamefont{Nordlander}},
  \bibinfo{journal}{{Nano Lett.}} \textbf{\bibinfo{volume}{{7}}},
  \bibinfo{pages}{729} (\bibinfo{year}{{2007}}).

\bibitem{Mish-book}
\bibinfo{author}{\bibfnamefont{M.~I.} \bibnamefont{Mishchenko}},
  \bibinfo{author}{\bibfnamefont{L.~D.} \bibnamefont{Travis}},
  \bibnamefont{and} \bibinfo{author}{\bibfnamefont{A.~A.} \bibnamefont{Lacis}},
  \emph{\bibinfo{title}{Scattering, Absorption and Emission of Light by Small
  Particles}} (\bibinfo{publisher}{Cambridge U. Press},
  \bibinfo{address}{Cambridge}, \bibinfo{year}{2002}).

\bibitem{J-C}
\bibinfo{author}{\bibfnamefont{P.~B.} \bibnamefont{Johnson}} \bibnamefont{and}
  \bibinfo{author}{\bibfnamefont{R.~W.} \bibnamefont{Christy}},
  \bibinfo{journal}{Phys. Rev. B} \textbf{\bibinfo{volume}{6}},
  \bibinfo{pages}{4370} (\bibinfo{year}{1972}).

\bibitem{JOSAA07}
\bibinfo{author}{\bibfnamefont{V.}~\bibnamefont{Giannini}} \bibnamefont{and}
  \bibinfo{author}{\bibfnamefont{J.~A.} \bibnamefont{S{\'a}nchez-Gil}},
  \bibinfo{journal}{J. Opt. Soc. Am.~ A} \textbf{\bibinfo{volume}{24}},
  \bibinfo{pages}{2822} (\bibinfo{year}{2007}).

\bibitem{PRB03spp}
\bibinfo{author}{\bibfnamefont{J.~A.} \bibnamefont{S{\'a}nchez-Gil}},
  \bibinfo{journal}{Phys. Rev. B} \textbf{\bibinfo{volume}{68}},
  \bibinfo{pages}{113410} (\bibinfo{year}{2003}).

\end{thebibliography}

\end{document}